\documentclass[aps,prl,twocolumn,superscriptaddress,longbibliography]{revtex4-2}

\usepackage{scalerel}
\usepackage{amssymb}
\usepackage{suffix}
\usepackage{float}
\usepackage{mathtools}
\usepackage[autostyle]{csquotes}
\usepackage[utf8]{inputenc}
\usepackage{booktabs}
\usepackage{cases}
\usepackage[multiple]{footmisc}
\usepackage{dcolumn}
\usepackage{color,soul}
\usepackage{rotating}
\usepackage{perpage}
\usepackage{xcolor}
\usepackage{soul}
\usepackage{tikz}
\usepackage[T1]{fontenc}
\usepackage{etoolbox}
\usepackage{graphics}
\usepackage{siunitx}
\usepackage[hidelinks]{hyperref}
\hypersetup{
	colorlinks = {true},
	linkcolor={red},
	citecolor={blue},
	urlcolor={blue}
}
\usepackage{float}	
\usepackage{collref}
\usepackage{multirow}
\usepackage{mathtools}
\usepackage{bm}
\usepackage{url}

\usepackage{tikz}
\usepackage{tikz-3dplot}
\usepackage{changes}

\usepackage{etoolbox,lipsum}

\newcommand{\blue}[1]{\textcolor{blue}{#1}}

\begin{document} 
	\title{Cavity-assisted magnetization switching in a quantum spin-phonon chain}
	
	\author{Mohsen Yarmohammadi}
	\email{mohsen.yarmohammadi@georgetown.edu}
	\address{Department of Physics, Georgetown University, Washington DC 20057, USA}
	\author{Peter M. Oppeneer}
	\address{Department of Physics and Astronomy, Uppsala University, Box 516, SE-75120 Uppsala, Sweden}
	\author{James K. Freericks}
	\address{Department of Physics, Georgetown University, Washington DC 20057, USA}

	\date{\today}
	
	\begin{abstract}
		We propose a Néel magnetization switching mechanism in a hybrid magnon–phonon optical cavity system. A terahertz-pumped single-mode cavity photon couples to a spin–phonon chain, while the system dissipates energy via different baths. Our mean-field analysis reveals that the photon induces magnetization switching by generating strongly entangled magnon pairs with opposite momenta—a feature weakly present in the cavity-free system. This switching occurs only at specific drive frequencies, namely at low magnon energies and near the upper edge~(perpendicular modes) of the two-magnon band. Our results underscore the roles of laser fluence, damping, and photon loss in modulating the switching process, offering a promising route for cavity-assisted magnetization control in opto-spintronics.
	\end{abstract}
	
	\maketitle
	{\allowdisplaybreaks		
		
		
		
		\blue{\textit{\textbf{Introduction}}}.---Spintronics exploits electron spin to enhance electronics, with ferromagnets traditionally favored for their ease of measurement and control~\cite{RevModPhys.76.323}. Recently, quantum antiferromagnets (AFMs) have emerged as promising alternatives for data storage and ultrafast processing due to their stray-field-free nature and inherent THz energy scale that enables picosecond switching~\cite{doi:10.1126/science.aad8211,https://doi.org/10.1002/pssr.201700022,Chappert2007,PhysRev.85.329,Jungwirth2016}. Realizing efficient AFM devices—and thus faster, more efficient technologies~\cite{PhysRevB.98.024422,PhysRevApplied.14.014004}—hinges on precise control of Néel magnetization to reliably encode binary information~\cite{Jungwirth2016, PhysRevB.109.174419,PRXQuantum.4.030332,PhysRevB.111.064408,PhysRevB.111.054411}.
		
		Ultrafast laser control of quantum materials reveals new phenomena~\cite{Rini2007,Forst2011,doi:10.1146/annurev-matsci-070813-113258,Mitrano2016,PhysRevB.94.214504,PhysRevLett.118.087002,PhysRevB.95.024304,Kennes2017,PhysRevB.95.205111,PhysRevLett.121.097402,Basov2017}. Advances in ultrafast spectroscopy enable efficient AFM order control without macroscopic magnetization and rapid manipulation via high energy scales~\cite{10.1063/1.2199473,doi:10.1126/science.aab1031,Nemec2018,Kampfrath2011,doi:10.1126/sciadv.aar3566,Li2020,doi:10.1126/science.aaz4247,10.1063/5.0075999,Mrudul2020,PhysRevB.109.144418,PhysRevB.108.064427,10.1063/1.4862467,Kampfrath2011,doi:10.1126/science.1214131}. Yet, classical lasers suffer from heating and cooldown delays. Embedding materials in optical cavities extends the lifetime of laser-induced states, reduces heating, and enhances light-matter coupling—especially in compact THz systems with significant quantum fluctuations~\cite{Ruggenthaler2018,FriskKockum2019,Hubener2021,10.1063/PT.3.4749,10.1063/5.0083825,Le_D__2022,Li2018,PhysRevB.99.235156,kipp2024cavityelectrodynamicsvander,PhysRevLett.131.023601,PhysRevB.105.205424,10.1063/5.0083825}. Systematic dissipation control is also crucial to balance energy injection and loss for engineered steady states~\cite{PhysRevB.108.L140305,PhysRevB.111.064414,PhysRevLett.134.050603}. Recently, one of us showed that linear and quadratic spin-phonon couplings~\cite{hart2024phonondriven,PhysRevLett.100.077201,Hernandez_2023,PhysRevB.94.014409} induce nonlinear magnetization dynamics in an AFM spin-phonon chain driven by a steady classical THz laser that excites infrared-active phonons~\cite{PhysRevB.110.134442}. Although these couplings trigger intriguing magnetic behaviors, they do not switch Néel magnetization~(arising from the magnetic moments of two adjacent sublattices)—a key for AFM opto-spintronics. \begin{figure}[t]
			\centering
			\includegraphics[width=0.85\linewidth]{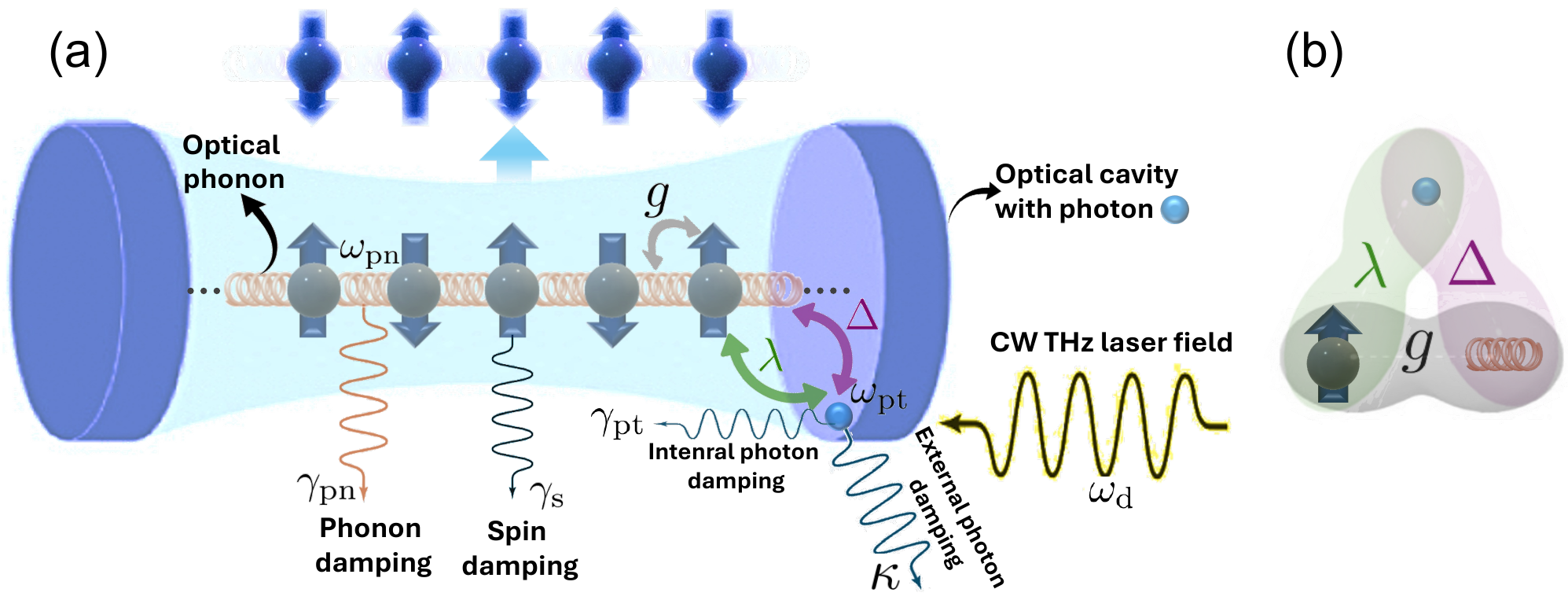}
			\caption{Spin-phonon chain in a terahertz laser‐driven cavity. (a) A spin chain (exchange \(J\), anisotropy \(\delta\)) interacts with a resonant cavity mode (\(\omega_{\rm pt}\)) driven by a continuous‐wave terahertz laser (\(\omega_{\rm d}\)). Infrared‐active optical phonons (\(\omega_{\rm pn}\)) couple to the spins with strength \(g\), while spin-photon (\(\lambda\)) and phonon-photon (\(\Delta\)) interactions hybridize states (b). Internal damping occurs at rates \(\gamma_{\rm pt}\), \(\gamma_{\rm pn}\), and \(\gamma_{\rm s}\), with additional photon loss (\(\kappa\)) to vacuum. The cross-talk between the driven-photon and spins triggers a magnetization switching, see the main text for details.}
			\label{f1}
		\end{figure}
		
		In this Letter, we demonstrate that an optical cavity can switch Néel magnetization in an AFM spin-phonon chain by generating magnon pairs with opposite momenta. We use a single-mode cavity photon with negligible energy density to minimally perturb spins and phonons. Although both couple to the photon mode, the weak spin-photon interaction~\cite{PhysRevB.101.205140,PhysRevResearch.2.033033,PhysRevB.103.075131,PhysRevB.99.085116} requires robust anisotropic systems—such as the Sm-Fe chain in SmFeO$_3$~\cite{PhysRevB.109.224417,Weber2022,PhysRevB.107.104427}—to overcome high switching activation energies~\cite{Song_2018,PhysRevB.109.174419,PRXQuantum.4.030332}. By weakly driving the cavity and including dissipation, we study how spin-photon and phonon-photon interactions affect the chain’s dynamics, ultimately achieving switched Néel magnetization via broken symmetry arising from lattice distortions due to  hybridization effects~\cite{PhysRevB.98.094412}.
		
		\blue{\textit{\textbf{Model}}}.---Figure~\ref{f1} illustrates our model. A terahertz-pumped single-mode cavity photon couples to a spin–phonon chain, with the system dissipating energy through various baths. The Hamiltonian model, outlined further below, comprises seven parts—the anisotropic XXZ spin chain, optical phonons, a single-mode cavity photon, spin-phonon, phonon-photon, and spin-photon couplings, plus the laser-cavity interaction—and we now introduce physically motivated approximations to simplify it.
		
		We adopt three approximations. First, the dipole approximation~\cite{PhysRevB.111.085114}—valid for the long wavelengths of optical cavities—neglects the spatial dependence of the electromagnetic vector potential, simplifying the photonic operators $\{b, b^\dagger\}$. Second, we focus on low-momentum optical phonon modes, justified by the scale disparity between the atomic lattice spacing and THz laser wavelengths, which introduces phononic operators $\{a, a^\dagger\}$. Third, a mean-field decoupling method constructs the coupling Hamiltonian between photons, spins, and phonons. This method redistributes laser energy among these components and is valid for long chains where quantum fluctuations are suppressed, as photon and phonon numbers scale with the chain length $L$.
		
		We focus on long-time nonequilibrium steady states (NESS) to capture persistent dynamics. Since pulsed lasers only probe transient phenomena, we use a steady laser field, \(\mathcal{E}(t)=\mathcal{E}_0\cos(\omega_{\rm d}t)\) (with amplitude \(\mathcal{E}_0\) and frequency \(\omega_{\rm d}\))~\cite{PhysRevLett.122.017401,Sentefeaau6969}. The laser polarization is aligned with photon motion to maximize the electric field–photon displacement coupling. With dipole (for photons) and low-momentum (for phonons) approximations, the phonon-photon interaction is modeled as two coupled harmonic oscillators~\cite{mahan2013many}.
		
		The full time-dependent Hamiltonian (with \(\hbar=1\)) is\begin{equation}\label{eq_1}
			\begin{aligned}
				&\mathcal{H}(t)={}J\sum_{\ell=1}^L\Big[\tfrac{1}{2}\left(S^+_\ell S^-_{\ell+1}+S^-_\ell S^+_{\ell+1}\right)+\delta S^z_\ell S^z_{\ell+1}\Big]\\
				&+\omega_{\rm pn}\,a^\dagger a+\omega_{\rm pt}\,b^\dagger b+\frac{g}{J}\frac{(a^\dagger+a)}{\sqrt{L}}\,\mathcal{H}_{\rm s}+\frac{\lambda}{J}\frac{(b^\dagger+b)}{\sqrt{L}}\,\mathcal{H}_{\rm s}\\
				&+\Delta\Big[\frac{2\Delta}{L\omega_{\rm pn}}(b^\dagger+b)^2-i\sqrt{L}(b^\dagger+b)(a^\dagger-a)\Big]\\
				&+\mathcal{E}(t)\sqrt{L}(b^\dagger+b)\,,
			\end{aligned}
		\end{equation}where the spin Hamiltonian, \(\mathcal{H}_{\rm s}\), refers to the first term, [\dots], and \(J\) (with \(\delta\simeq1.2\)) is the spin-spin coupling; \(\omega_{\rm pn}\) and \(\omega_{\rm pt}\) are the phonon and photon frequencies; and \(S\), \(a\), and \(b\) denote the spin, phonon, and photon operators. Here, \(g\) and \(\lambda\) quantify the spin-phonon and spin-photon couplings, while \(\Delta=\sqrt{\omega_{\rm p}\omega_{\rm pn}/8}\)~($\omega_{\rm p}$ being the polariton frequency) sets the phonon-photon interaction strength with \(0<\omega_{\rm p}/\omega_{\rm pt}<0.005\)~\cite{Le_D__2022,Eckhardt2022,mahan2013many}.
		
		In the spin sector (Fig.~\ref{f1}), we consider an anisotropic AFM chain with \(S>1/2\) and nearest-neighbor Heisenberg interactions (\(S^\pm=S^x\pm iS^y\)) between two sublattices. Approximating spins as classical vectors, we rotate the axis on the down sublattice to align with the Néel state~\cite{PhysRevB.109.224417}. Using the Holstein-Primakoff transformation~\cite{PhysRev.58.1098}, we write, for example, \(S^z_\ell = m^\dagger_\ell m_\ell - S\) and \(S^+_\ell \simeq \sqrt{2S}\,m_\ell\) (with similar expressions for the down sublattice at $\ell + 1$). Diagonalizing the spin Hamiltonian via a Bogoliubov transformation, $m^\dagger_k=\cosh(\theta_k)\,\widetilde{m}^\dagger_k+\sinh(\theta_k)\,\widetilde{m}_{-k}$, with \(\sinh(2\theta_k)=-\cos(k)/\sqrt{\delta^2-\cos^2(k)}\), yields the magnon dispersion $\omega_k = 2 J S \sum_k \sqrt{\delta^2 - \cos^2(k)}$.
		
		In the spin-phonon sector, we introduce a local interaction at each site, reflecting the structure of real materials—two magnetic ions bridged by an intermediate charged ion. Driven photons induce phonon excitations that create disorder, and the resulting vibrations of the magnetic ions modulate the exchange coupling at low temperatures. We model this by expanding the phonon-dependent exchange couplings, so that the relative sublattice oscillations drive the spin-phonon interaction~\cite{PhysRevB.109.224417}. Similarly, in the spin-photon coupling, the photon directly induces ionic vibrations that modulate the exchange coupling, allowing us to define an averaged magnon mode \(k\) for coupling to both photons and phonons.
		
		We model the spin-phonon and spin-photon couplings by incorporating Hamiltonian terms from both easy-plane and easy-axis spin interactions, then decoupling excitations via a mean-field approximation~\cite{PhysRevB.109.224417, yarmohammadi2020dynamical, allafi2024spin, PhysRevB.107.174415, yarmohammadi2024ultrafast, PhysRevB.110.134442, PhysRevB.108.L140305}:\begin{equation}
			\small \begin{aligned}
				\mathcal{H}^{\rm MF}_{\rm spn+spt}           \approx {} & \left[\frac{g}{\sqrt{L}} \langle a^\dagger + a\rangle + \frac{\lambda}{\sqrt{L}} \langle b^\dagger + b\rangle\right] \sum_k \frac{2 J S \cos(k)}{\sqrt{\delta^2 - \cos^2(k)}}\\
				{} & \times \left(\delta\, \widetilde{m}^\dagger_k \widetilde{m}_k - \widetilde{m}^\dagger_{k} \widetilde{m}^\dagger_{-k}\right)\, .
			\end{aligned}
		\end{equation}
		
		Lastly, we capture dissipation by coupling our system to two baths: an internal bosonic bath (accounting for undriven cavity photon modes and uncoupled phonons) and an external vacuum bath (modeling photon loss). The evolution of any observable $O$ is given by the Lindblad master equation \cite{lindblad1976,breuer2007theory}:\begin{equation}\label{eq_5} 
			\langle \dot{O}\rangle (t) = i\langle[\mathcal{H}(t),O]\rangle  + \sum_{i} \gamma_{i} \big<\mathcal{L}_{i}^{\dagger}O\mathcal{L}_{i}  -\frac{1}{2}\{\mathcal{L}_{i}^{\dagger}\mathcal{L}_{i},O\}\big>(t), 
		\end{equation}with jump operators $\mathcal{L}_{i}$ chosen from $\{a,b,\widetilde{m}_k\}$ for the photon, phonon, and magnon subsystems, respectively. Here, the damping rates $\gamma_i$ correspond to the cavity photons~\((\gamma_{\rm pt})\), phonons~\((\gamma_{\rm pn})\), spins~(\(\gamma_{\rm s}\)), and to external photon loss~($\kappa$).\begin{figure*}[t]
			\centering	\includegraphics[width=0.85\linewidth]{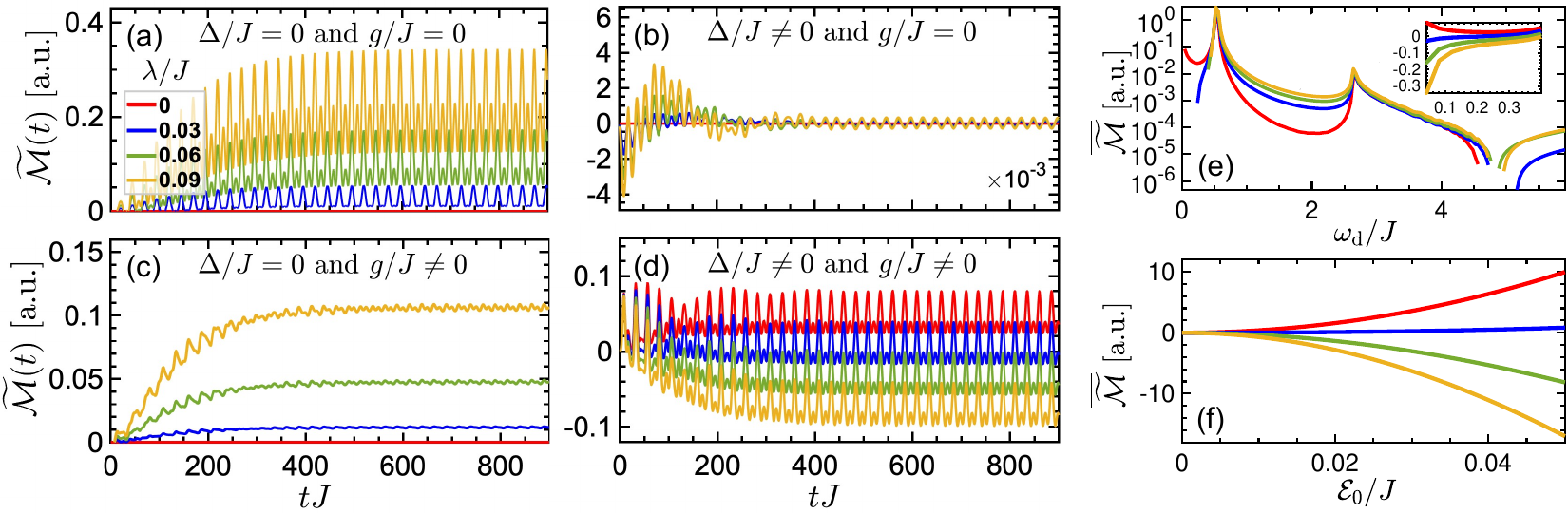}
			\caption{Time evolution of Néel magnetization for different coupling scenarios: (a) $\Delta=0$, $g=0$; (b) $\Delta=0.0028\,J$, $g=0$; (c) $\Delta=0$, $g=J$; and (d) $\Delta=0.0028\,J$, $g=J$, with spin–photon coupling $0<\lambda/J<0.1$. Only when both spin–photon ($\lambda$) and phonon–photon ($\Delta$) interactions are active [panel (d)] does switching/demagnetization occur. Fixed parameters: $\omega_{\rm d}=\omega_{\rm pt}=\omega_{\rm pn}=J/4$, $\gamma_{\rm pn}=\gamma_{\rm pt}=\kappa=J/80$, $\mathcal{E}_0=J/400$, and $\gamma_{\rm s}=J/100$. Laser effects on switching are shown in (e) (frequency) and (f) (fluence): low drive frequencies (reflecting the low-energy states at low temperatures) and drives near the upper two-magnon band edge switch magnetization~(negative values) with a power-law fluence dependence. For (e) and (f): $\Delta=0.0028\,J$, $g=J$, $\omega_{\rm pt}=\omega_{\rm pn}=J/4$, $\gamma_{\rm pn}=\gamma_{\rm pt}=\kappa=J/80$, and $\gamma_{\rm s}=J/100$; $\mathcal{E}_0=J/400$ in (e) and $\omega_{\rm d}=J/4$ in (f).} 
			\label{f2}
		\end{figure*}  
		
		For the observables defined in the Appendix \hyperlink{mylinkA}{A}, simplification yields equations of motion for all particles at low temperatures with equilibrium initial conditions \(O(0)=0\). We numerically run these equations sufficiently long to attain a true NESS. The Néel magnetization per spin is given by $\mathcal{M}(t)\approx\frac{1}{L}\sum_{\ell}\Big[\langle S^z_\ell\rangle(t)-\langle S^z_{\ell+1}\rangle(t)\Big]$. Defining \(\widetilde{\mathcal{M}}(t)=\mathcal{M}(t)/(2JS)\) leads to
		\begin{equation}\label{eq_4}
			\widetilde{\mathcal{M}}(t)=\frac{1}{L}\sum_k\frac{1}{\sqrt{\delta^2-\cos^2(k)}}\Big[\delta\,n_{{\rm s},k}(t)-\cos(k)\,v_{{\rm s},k}(t)\Big],
		\end{equation}
		where $n_{{\rm s},k}(t)=\langle\widetilde{m}^\dagger_{k}\widetilde{m}_{k}\rangle(t)$ and $v_{{\rm s},k}(t)=\Re[\langle\widetilde{m}^\dagger_{k}\widetilde{m}^\dagger_{-k}\rangle(t)]$ represent the \(k\)-components of the magnon density and off-diagonal spin excitations (magnon pairs), respectively.
		
		We compute the Néel magnetization dynamics by numerically solving Eqs.~\eqref{eq_s6}–\eqref{eq_s7} in the Appendix \hyperlink{mylinkA}{A} for an \(L=2001\) chain to eliminate finite-size effects. At low temperatures, with such a large system, magnon excitations are sparse. The resulting spin excitations \(v_{{\rm s},k}(t)\) and magnon density \(n_{{\rm s},k}(t)\) are then substituted into Eq.~\eqref{eq_4}. Since negative spin density is unphysical, the switching process is predominantly driven by the generation of magnon pairs with opposite magnetic moments. Our weak-coupling analytical expression, as detailed in the Appendix \hyperlink{mylinkB}{B}---valid when $\xi_k < \gamma_{\rm s}$, where $\xi_k = 4J_1\big(2\tfrac{\delta}{\omega_{\rm d}}(g q^1_{\rm pn} + \lambda q^1_{\rm pt})\big)(gq^1_{\rm pn} + \lambda q^1_{\rm pt})\cos(k) \omega_{\rm d}/\delta$ with Bessel function $J_1(\dots)$, $q^1_{\rm pn} \approx 2 \mathcal{E}_0 + \Delta/2$ and $q^1_{\rm pt} \approx 2 \mathcal{E}_0 - \Delta/2$---proves that the dominant contribution from magnon pairs with opposite moments, \(v_{{\rm s},k}(t)\), can drive the Néel magnetization negative under certain coupling conditions. This, in turn, highlights the significance of spin damping relative to other types of damping.
		
		\blue{\textit{\textbf{Results and discussion}}}.---In establishing a physically relevant parameter space, the laser amplitude is first selected to ensure system stability while preventing melting and decoherence. Photon damping is then set to approximately a few percent of the photon energy, consistent with the single-mode photon and dipole approximation. A similar approach is applied to the phonon, based on its low-momentum approximation. Moreover, the spin damping is maintained below both the phonon and photon damping, reflecting the inherently weak spin interactions. Finally, the spin–phonon and phonon–photon couplings are restricted to the weak coupling regime to ensure experimental feasibility. These choices guarantee that the combined effects of coupling and damping effectively yield a genuine NESS. All parameters are expressed in units of $J=1$, corresponding to an energy scale on the order of THz.
		
		We concentrate on resonant conditions—specifically, setting \(\omega_{\rm d} = \omega_{\rm pt} = \omega_{\rm pn}\)—to elicit the strongest responses in the system. Notably, our findings remain consistent for off-resonant drive frequencies below the threshold \(\omega_{\rm pt} = \omega_{\rm pn}\). In contrast, for frequencies exceeding this threshold, the switching process is effectively suppressed (see Appendix \hyperlink{mylinkC}{C}).
		
		A central parameter in our study is the spin–photon coupling, \(\lambda\), which underpins our investigation into the orientation and magnitude of the Néel magnetization in the presence of a single-mode cavity photon. To systematically delineate its impact, we vary \(\lambda\) while monitoring the influence of all other model parameters. Given that magnetism is inherently weak, we constrain \(\lambda\) to remain below 10\% of the bare magnetic exchange coupling, \(J\). This upper bound is validated by our numerical simulations, where \(\lambda\) consistently emerges alongside the other parameters in a phenomenologically coherent manner.
		
		Based on the parameter framework established above, we now present the ultrafast dynamics of the Néel magnetization in the hybrid magnon–phonon optical cavity. Our analysis reveals that the system attains a NESS on a timescale of approximately 66 picoseconds for \(J = 1\) THz. Furthermore, by selecting appropriate parameters, a genuine long-time NESS is achieved.
		
		In Fig.~\ref{f2}, we emphasize the roles of the couplings \(\lambda\), \(g\), and \(\Delta\) in realizing our central objectives. In Fig.~\ref{f2}(a), with phonon degrees of freedom absent (\(g=\Delta=0\)), the spin couples directly to the driven photon, and an increase in the spin–photon coupling \(\lambda\) enhances the magnetization. This is attributed to \(\lambda\) acting as a perturbative element that increases magnon occupation, in agreement with perturbation theory, which in turn amplifies the magnetization. In contrast, Fig.~\ref{f2}(b) considers the case where spin–phonon coupling is absent (\(g=0\)) while phonon–photon coupling is present (\(\Delta\neq0\)). Here, the spin–photon interaction immediately initiates Néel magnetization switching, as the spin excitations—represented by magnon pairs \(v_{{\rm s},k}(t)\)—are significantly affected by both \(\lambda\) and \(\Delta\), leading to \(\widetilde{\mathcal{M}}(t)<0\) in Eq.~\eqref{eq_4}. However, this switching behavior is confined to the transient dynamics, with no persistent dominant oscillations observed in the long-time NESS.
		
		In Fig.~\ref{f2}(c), where phonon–photon coupling is absent (\(\Delta=0\)) but spin–phonon coupling is present (\(g\neq0\)), the Néel magnetization increases with the spin–photon coupling \(\lambda\). This trend can be attributed to the perturbative effect of \( g \) on the system, which effectively adds to the role of \( \lambda \). In contrast, Fig.~\ref{f2}(d) illustrates that when both spin–phonon and phonon–photon couplings are active (\(\Delta\neq0\) and \(g\neq0\)), the system achieves its primary objective of magnetization switching and demagnetization. Here, the phonon–photon interaction amplifies the magnon pair excitations to a degree that drives \(\widetilde{\mathcal{M}}(t)\) below zero~($\lambda /J > 0.03$). These results underscore the necessity of concurrently employing both phonon–photon and spin–photon couplings to realize Néel magnetization switching and demagnetization in the spin chain.
		
		Next, we assess the influence of additional experimentally relevant parameters on magnetization switching and demagnetization by evaluating the time-averaged Néel magnetization in the NESS, defined over one oscillation period as $\overline{\widetilde{\mathcal{M}}} = {} \tfrac{1}{T}\int_0^T \widetilde{\mathcal{M}}(t) d t$.
		
		We now examine the impact of laser driving parameters—specifically, the drive frequency and fluence—on the magnetization switching mechanism. With the anisotropy fixed at \(\delta = 1.2\), the one-magnon band spans approximately from \(1.325\,J\) to \(2.4\,J\) as the magnon mode evolves from \(k = 0\) to \(k = \pi/2\), returning to \(1.325\,J\) at \(k = \pi\). Accordingly, the two-magnon band is estimated to lie between \(2.65\,J\) and \(4.8\,J\). The Hamiltonian gives rise to two distinct resonances, see Fig. \ref{f2}(e), because both the photon and phonon couple to two spins within the chain. The first resonance occurs when \(\omega_{\rm d} = \omega_{\rm pt} = \omega_{\rm pn} = J/4\), where the driving frequency coincides with the characteristic frequencies of the phonon and photon modes. A second resonance is observed at $\omega_{\rm d} = 4 J S \sqrt{\delta^2 - \cos^2(k)}$, which emerges within the two-magnon band due to the spin–photon coupling. This interplay among the photon, phonon, and spin degrees of freedom produces the intricate resonance structure dictated by the Hamiltonian. From this point, as shown in Fig.~\ref{f2}(e), the drive frequency exhibits distinct features—a peak at the lower edge and a kink at the upper edge of the two-magnon band—irrespective of the spin–photon coupling \(\lambda\). 
		
		Notably, magnetization becomes negative in the upper two-magnon band, as evidenced by the absence of data points in the logarithmic plot. Furthermore, for finite \(\lambda\), low drive frequencies (see inset of Fig.~\ref{f2}(e)) also result in negative Néel magnetization, which is consistent with the condensation phase of the magnon–phonon cavity system, wherein hybridized states reside at lower energy levels.
		
		Figure~\ref{f2}(f) demonstrates that increasing the laser amplitude, $\mathcal{E}_0$, leads to a pronounced enhancement of the Néel magnetization for $\lambda/J = 0$ and $0.03$, while for $\lambda/J = 0.06$ and $0.09$, clear signatures of switching and demagnetization are observed. This behavior is consistent with Fermi's golden rule, which predicts that the system's response to an external driving field is modulated by the field strength. Within a perturbative framework, the magnetization exhibits a power-law scaling proportional to $\pm\,\mathcal{E}_0^2$. Moreover, to ensure the physical stability of the system under intense laser irradiation, the laser amplitudes were selected following the Lindemann criterion~\cite{Lindemann1910}, thereby preventing lattice melting or other structural instabilities.\begin{figure}[b]
			\centering
			\includegraphics[width=0.76\linewidth]{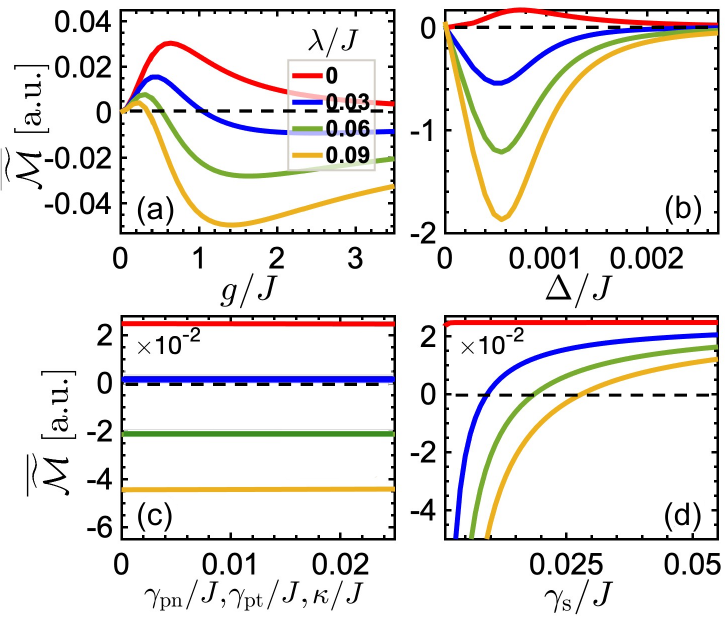}
			\caption{Dependence of averaged Néel magnetization on spin-phonon and phonon-photon couplings and damping parameters in a hybrid magnon–phonon optical cavity. In the weak-coupling regime, the response scales as \(g^2\) or \(\Delta^2\), with nonlinear effects emerging at higher values; increasing spin–photon coupling lowers the critical spin–phonon threshold for switching. Panels (a,b): fixed \(\Delta=0.0028J\) and \(g=J\); panels (c,d): photon/phonon damping leaves the response unchanged, while spin damping shows distinct critical values. Parameters: $\mathcal{E}_0=J/400$, $\omega_{\rm d}=\omega_{\rm pn}=\omega_{\rm pt}=J/4$, $\gamma_{\rm pn}=\gamma_{\rm pt}=\kappa=J/80$, and $\gamma_{\rm s}=J/100$.} 
			\label{f3}
		\end{figure}
		
		Figures~\ref{f3}(a) and \ref{f3}(b) illustrate the magnetization evolution as a function of coupling strengths, with the drive, photon, and phonon maintained in resonance. Per perturbation theory, the leading corrections to the system's properties scale quadratically with the perturbative parameters, namely the spin-phonon coupling, $g$, and the phonon-photon coupling, $\Delta$. At low values of $g$ and $\Delta$, the magnetization displays a clear $g^2$ and $\Delta^2$ dependence. However, as these couplings increase, a suppression of $\overline{\widetilde{\mathcal{M}}}$ emerges, signaling the onset of nonlinear effects. Although a revival of magnetization might be anticipated at very strong couplings, such a regime is improbable in most magnetic quantum materials. Furthermore, our results reveal that stronger spin-photon coupling reduces the critical spin-phonon coupling required for magnetization switching, underscoring the pivotal role of the spin-photon interaction in governing the final magnetization state.
		
		In our final analysis, we incorporate damping across spins, photons, and phonons—a crucial consideration in this driven-dissipative regime. As indicated by Eqs.~\eqref{eq_17} and \eqref{eq_18} in the Appendix \hyperlink{mylinkB}{B}, the effective damping governing the steady-state Néel magnetization is predominantly determined by the spin damping, which competes with the various coupling and driving parameters. This behavior can be ascribed to the cooperativity among particles~\cite{doi:10.1126/sciadv.1501286}. To quantify coherent interactions in a dissipative environment, high cooperativity during the magnetization switching process is essential. 
		
		We define the cooperativity as $C_{i,j}={}G_{i,j}^2/\gamma_i\,\gamma_j$,
		with \(G_{i,j}=\{g,\Delta,\lambda\}\) and $\gamma_{i/j}$ the corresponding damping rates~\cite{PhysRevLett.134.056701}. Our analysis reveals that, among the different damping channels, spin damping is chiefly responsible for achieving high cooperativity in coherent system interactions during the switching process. For our hybrid magnon-phonon cavity system in the switched phase, we find $C_{\rm s-pt}\approx14.4,\quad C_{\rm s-pn}\approx1.25,\quad \text{and}\quad C_{\rm pn-pt}\approx0.025$. These numbers, in turn, underscore the significance of the cavity photon being coupled to the spin. Accordingly, variations in $\gamma_{\rm pn}$, $\gamma_{\rm pt}$, and $\kappa$ have little impact on the NESS response (Fig.~\ref{f3}(c)). In contrast, Fig.~\ref{f3}(d) reveals that the critical spin damping is sensitive to the spin-photon coupling strength. 
		
		\blue{\textbf{\textit{Conclusions}}}.---Our results demonstrate a mechanism for magnetization switching and demagnetization in a hybrid magnon–phonon optical cavity system, addressing the longstanding challenge of magnetization control in antiferromagnetic spintronics. In our configuration, a spin–phonon chain is coupled to a single-mode optical cavity, where a terahertz pump generates the cavity photons. While previous studies have shown that spin–phonon coupling alone is insufficient to manipulate magnetization in gapped quantum magnets, our mean-field analysis reveals that the combined effects of magnon–phonon, phonon–photon, and magnon–photon interactions are critical for switching magnetization dynamics. Notably, the generation of entangled magnon pairs with opposite momenta, which occurs under specific drive conditions—either at low frequencies or near the upper edge of the two-magnon band—plays a central role in the switching process. 
		
		Furthermore, we detail how the magnetization response is influenced by experimentally tunable parameters, including laser fluence, damping rates, and photon loss. Our analysis reveals that photon and phonon damping exert minimal influence on the magnetization response, while spin damping is decisive, with its critical threshold values governing the switching behavior. This pronounced sensitivity to spin damping underscores the necessity of optimizing baths to achieve efficient and robust magnetization control. By elucidating the complex interplay among spins, phonons, and photons in this hybrid system, our findings offer a promising avenue for the advancement of opto-spintronic technologies.
		
		\blue{\textbf{\textit{Acknowledgments}}}.---M.Y. acknowledges the hospitality of Uppsala University during his visit. M. Y. and J.K.F. were supported by the Department of Energy, Office of Basic Energy Sciences, Division of Materials Sciences and Engineering under Contract No. DE-FG02-08ER46542 for the formal developments, the numerical work, and the writing of the manuscript. P.M.O. acknowledges support by the Swedish Research Council (VR), the German Research Foundation (Deutsche Forschungsgemeinschaft) through CRC/TRR 227 “Ultrafast Spin Dynamics” (project MF, project-ID: 328545488), and the K. and A. Wallenberg Foundation (Grants No. 2022.0079 and 2023.0336).  
	}
	
	\bibliography{bib_new.bib}
	{\allowdisplaybreaks
		\onecolumngrid
		\subsection{\large End Matter}\label{apa}
		\twocolumngrid
		\hypertarget{mylinkA}{\textbf{\blue{\textbf{\textit{Appendix A: Equations of motion}}}}}.---In this Appendix, we provide the equations of motion necessary to monitor the time evolution of the Néel magnetization. We begin our analysis by considering the Hamiltonian given in Eq.~\eqref{eq_1}. Our first step is to identify the most relevant physical observables in the model. 
		
		For the photon subsystem, we define the following observables following the Hamiltonian:\begin{subequations}\label{eq_2}
			\begin{align}
				q_{\rm pt}(t) = {}&  \big \langle \tfrac{1}{\sqrt{L}} (b^\dagger  + b) \big \rangle (t)\,,\\
				p_{\rm pt}(t) = {} &\big \langle \tfrac{i}{\sqrt{L}} (b^\dagger  - b)  \big \rangle (t)\, ,\label{eq_2a}\\
				\mathcal{Q}_{\rm pt}(t) = {} &  \big \langle \tfrac{1}{L} (b^\dagger b^\dagger  + bb) \big\rangle (t)\,, \\
				\mathcal{P}_{\rm pt}(t) =  {} & \big \langle \tfrac{i}{L} (b^\dagger b^\dagger-bb) \big\rangle (t)\, ,\label{eq_2d}
				\\
				n_{\rm pt}(t) = {} & \big \langle \tfrac{1}{L} b^\dagger b   \big \rangle (t)\,.\label{eq_2e}
			\end{align}
		\end{subequations}Similarly, the phononic observables can be defined by replacing the subscript \enquote{pt} with \enquote{pn} and substituting the operator \enquote{$b$} with \enquote{$a$} only for the position, momentum, and occupation observables. In both subsystems, $q$ and $p$ represent displacement and momentum, respectively. The quantities $\mathcal{Q}$ and $\mathcal{P}$ correspond to the squeezing parameters, while $n$ denotes the photon or phonon number in the respective subsystem. 
		
		For the spin sector, we introduce\begin{subequations}
			\begin{align}
				n_{{\rm s},k}(t) = {} & \big \langle  \widetilde{m}^\dagger_{k}  \widetilde{m}_{k}   \big \rangle (t)  \, ,\\
				z_{{\rm s},k}(t) = {} & \big \langle  \widetilde{m}^\dagger_{k}  \widetilde{m}^\dagger_{-k}   \big \rangle (t)  \, ,
			\end{align}
		\end{subequations}representing the $k$-component of the magnon density and the $k$-component of off-diagonal spin excitations (magnon pairs), respectively.
		
		Employing the Lindblad quantum master equation for the above observables, $\langle \dot{O}\rangle (t) = i\langle[\mathcal{H}(t),O]\rangle  + \sum_{i} \gamma_{i} \big<\mathcal{L}_{i}^{\dagger}O\mathcal{L}_{i}  -\frac{1}{2}\{\mathcal{L}_{i}^{\dagger}\mathcal{L}_{i},O\}\big>(t)$, as well as applying the Maxwell-Boltzmann factor to the ratio of damping between two dissipators $\mathcal{L}_{i} = O$ and $O^\dagger$, we obtain the following relations: $\gamma^{\rm s}_{2}/\gamma^{\rm s}_1 = \gamma_{\rm s}\mathcal{N}_k(t)/[1+\mathcal{N}_k(t)]$, $\gamma^{\rm pn}_{2}/\gamma^{\rm pn}_1 = \gamma_{\rm pn}\mathcal{N}_0(t)/[1+\mathcal{N}_0(t)]$, $\gamma^{\rm pt}_{2}/\gamma^{\rm pt}_1 = \gamma_{\rm pt} \mathcal{N}(t)/[1+\mathcal{N}(t)]$ and $\kappa_2/\kappa_1 = \kappa \mathcal{N}(t)/[1+\mathcal{N}(t)]$, where $\mathcal{N}(t)$($\mathcal{N}_0(t)$) represents the mean number of energy quanta in the photon (phonon) mode at time $t$, while $\mathcal{N}_k(t)$ corresponds to the mean number of energy quanta in the magnon mode $k$~\cite{lindblad1976,breuer2007theory}. Ultimately, we achieve\begin{widetext}\begin{subequations} \label{eq_s6} 
				\begin{align}  
					\dot{q}_{\rm pt}(t) = {} & \omega_{\rm pt} \,p_{\rm pt}(t)- \frac{\gamma_{\rm pt} + \kappa}{2}q_{\rm pt}(t)\, ,\label{eq_s6a}\\
					\dot{p}_{\rm pt}(t) = {} & -\left(\omega_{\rm pt} + \tfrac{8\Delta^2}{L\omega_{\rm pn}}\right)\,q_{\rm pt}(t) - 2\mathcal{E}(t) -2\lambda \mathcal{C}_{\rm s}(t) + 2 \Delta \sqrt{L}\,p_{\rm pn}(t) -  \frac{\gamma_{\rm pt}+ \kappa}{2}p_{\rm pt}(t)\, ,\label{eq_s6b}\\
					\dot{n}_{\rm pt}(t) = {} & - \left[\mathcal{E}(t)+\lambda \mathcal{C}_{\rm s}(t) -  \Delta \sqrt{L}\, p_{\rm pn}(t) \right]p_{\rm pt}(t)  - \tfrac{4\Delta^2}{L \omega_{\rm pn}}\mathcal{P}_{\rm pt}(t) - (\gamma_{\rm pt}+ \kappa) n_{\rm pt}(t)\, ,\label{eq_s6c}\\
					\dot{\mathcal{Q}}_{\rm pt}(t) = {} & \Big[2\omega_{\rm pt} + \tfrac{4\Delta^2}{L\omega_{\rm pn}}\Big]\mathcal{P}_{\rm pt}(t) + \Big[2\mathcal{E}(t) + 2\lambda \mathcal{C}_{\rm s}(t) - 2 \Delta \sqrt{L} p_{\rm pn}(t)\Big] p_{\rm pt}(t) - (\gamma_{\rm pt}+ \kappa) \mathcal{Q}_{\rm pt}(t)\, ,\label{eq_s6d}\\
					\dot{\mathcal{P}}_{\rm pt}(t) = {} & - \Big[2\omega_{\rm pt} - \tfrac{8\Delta^2}{L\omega_{\rm pn}}\Big]\mathcal{Q}_{\rm pt}(t) - \Big[2\mathcal{E}(t) + 2\lambda \mathcal{C}_{\rm s}(t) - 2 \Delta \sqrt{L} p_{\rm pn}(t)\Big] q_{\rm pt}(t) - \tfrac{8 \Delta^2}{L\omega_{\rm pn}}\left(2 n_{\rm pt}(t) + \tfrac{1}{L}\right)- (\gamma_{\rm pt}+ \kappa) \mathcal{P}_{\rm pt}(t)\, ,\label{eq_s6e}
				\end{align} 
		\end{subequations}\end{widetext}where $\mathcal{C}_{\rm s}(t) = {} \frac{1}{L} \sum_k \left[\delta\, n_{{\rm s},k}(t) + 2 \cos(k) v_{{\rm s},k}(t)\right]$. For the spin, we have 
		\begin{widetext}\begin{subequations}\label{eq_s11}
				\begin{align}
					\dot{n}_{{\rm s},k}(t) = {} & 2 \left[g q_{\rm pn}(t) + \lambda q_{\rm pt}(t)\right] \cos(k) w_{{\rm s},k}(t) - \gamma_{\rm s} n_{{\rm s},k}(t)\, ,\label{eq_9a}\\
					\dot{v}_{{\rm s},k}(t) = {} & -2 \Big(2 J S \sqrt{\delta^2 - \cos^2(k)} + \delta \,\big[g q_{\rm pn}(t) + \lambda q_{\rm pt}(t)\big]\Big)w_{{\rm s},k}(t) - \gamma_{\rm s} v_{{\rm s},k}(t)\, ,\label{eq_11b}\\
					\dot{w}_{{\rm s},k}(t) = {} & 2 \Big(2 J S \sqrt{\delta^2 - \cos^2(k)} + \delta \,\big[g q_{\rm pn}(t) + \lambda q_{\rm pt}(t)\big]\Big)v_{{\rm s},k}(t) + 4 \left(g q_{\rm pn}(t) + \lambda q_{\rm pt}(t)\right)  \cos(k) \left(n_{{\rm s},k}(t) + \tfrac{1}{2}\right)- \gamma_{\rm s} w_{{\rm s},k}(t)\, ,\label{eq_11c}
				\end{align} 
		\end{subequations}\end{widetext}and finally, for the phonon, we obtain
		\begin{subequations} \label{eq_s7} 
			\small{\begin{align}
					\dot{q}_{\rm pn}(t) = {} & \omega_{\rm pn} \,p_{\rm pn}(t) - 2 \Delta \sqrt{L} \,q_{\rm pt}(t)-   \frac{\gamma_{\rm pn}}{2}q_{\rm pn}(t)\, ,\label{eq_s7f}\\ 
					\dot{p}_{\rm pn}(t) =  {} &- \omega_{\rm pn} \, q_{\rm pn}(t) - 2 g \,\mathcal{C}_{\rm s}(t)- \frac{\gamma_{\rm pn}}{2} p_{\rm pn}(t) ,\label{eq_s7g}\\
					\dot{n}_{\rm pn}(t) = {} & - g \,\mathcal{C}_{\rm s}(t) \, p_{\rm pn}(t) - \Delta \sqrt{L} \,q_{\rm pt}(t)\,q_{\rm pn}(t)- \gamma_{\rm pn} n_{\rm pn}(t)\, .\label{eq_s7h}
			\end{align} }
		\end{subequations}
		
		\hypertarget{mylinkB}{\blue{\textbf{\textit{Appendix B: Analytical solution in the weak-coupling regime}}}}.---In this Appendix, we derive an analytical expression for the magnon density and magnon pairs necessary to establish Néel magnetization. However, this derivation is limited to the weak coupling regime between all three particles, where approximations can be employed to obtain exact analytical results. In the weak coupling regime, a cosine signal can be used in the NESS for photon and phonon displacements, so the excited photon or phonon behaves like a driven damped harmonic oscillator. This approach has also been numerically validated across a wide range of parameters (not shown here). Next, we focus exclusively on the resonance condition, where $\omega_{\rm d} = \omega_{\rm pt} = \omega_{\rm pn} = 4JS\sqrt{\delta^2 - \cos^2(k)}$. Under this condition, the magnon frequency is doubled, as a single phonon or photon mode couples to two spins on sites $\ell$ and $\ell+1$. Accordingly, we adopt the expressions $q_{\rm pn}(t) = q^1_{\rm pn} \cos(\omega_{\rm d} t)$ and $q_{\rm pt}(t) = q^1_{\rm pt} \cos(\omega_{\rm d} t)$, where \begin{align}
			q^1_{\rm pn} \approx 2 \mathcal{E}_0 + \Delta/2\quad,\quad q^1_{\rm pt} \approx 2 \mathcal{E}_0 - \Delta/2\, ,
		\end{align}represent the validated oscillation amplitudes numerically due to polaritonic splitting effect. Combining Eqs.~\eqref{eq_11b} and~\eqref{eq_11c} into a single equation for\begin{align}
			z_{{\rm s},k}(t) = v_{{\rm s},k}(t) + i w_{{\rm s},k}(t)\, ,
		\end{align}we obtain $\dot{z}_{{\rm s},k}(t) = [i \dot{h}_k(t) - \gamma_{\rm s}] z_{{\rm s},k}(t) + i f_k(t)$ with\begin{subequations}
			\begin{align}
				h_k(t) =&4JS\sqrt{\delta^2 - \cos^2(k)} t+ 2\tfrac{\delta}{\omega_{\rm d}}[g q^1_{\rm pn} + \lambda q^1_{\rm pt}]\sin(\omega_{\rm d}t) ,\\
				f_k(t) = & 4[g q^1_{\rm pn} + \lambda q^1_{\rm pt}]\cos(\omega_{\rm d}t)(n_{{\rm s},k}(t)+1/2)\,.
			\end{align}
		\end{subequations}To simplify the expressions, we focus on the slowly varying component of the magnon density, $n_{{\rm s},k}(t)$, and average it over one period, $T_{\rm d} = 2\pi/\omega_{\rm d}$, in the integrals. This leads to\small \begin{equation}\label{eq_15}
			\begin{aligned}
				&z_{{\rm s},k}(t) =  4i(g q^1_{\rm pn} + \lambda q^1_{\rm pt})\cos(k) e^{ih_k(t) - \gamma_{\rm s}t}
				\\ {} &\hspace{-0.4cm}\int^t_0 \Big[\frac{1}{T_{\rm d}} \int^{T_{\rm d}}_0 \cos(\omega_{\rm d}t'') e^{-i\omega_{\rm d}t''- 2i\tfrac{\delta}{\omega_{\rm d}}(g q^1_{\rm pn} + \lambda q^1_{\rm pt})  \sin(\omega_{\rm d}t'')} d t''\Big]\\ {} & (n_{{\rm s},k}(t) + 1/2)e^{\gamma_{\rm s}t'} dt'\, ,\\
				= {} & \tfrac{4 i (gq^1_{\rm pn} + \lambda q^1_{\rm pt}) \cos(k) \omega_{\rm d}}{\delta} J_1\big(2\tfrac{\delta}{\omega_{\rm d}}(g q^1_{\rm pn} + \lambda q^1_{\rm pt})\big) e^{ih_k(t) - \gamma_{\rm s}t} \\ {} &\int^t_0 (n_{{\rm s},k}(t) + 1/2)e^{\gamma_{\rm s}t'} dt'\, ,
			\end{aligned}
		\end{equation}where $J_1(\dots)$ is the Bessel function of the first kind. Using the relation $w_{{\rm s},k}(t) = -(i/2)[z_{{\rm s},k}(t) - z^*_{{\rm s},k}(t)]$ in Eq.~\eqref{eq_11c}, we find\begin{equation}
			n_{{\rm s},k}(t) = \xi_k^2 e^{-\gamma_{\rm s}t} \int^t_0 \int^{t'}_0 (n_{{\rm s},k}(t'') + 1/2)e^{\gamma_{\rm s}t''} dt''d t'\, ,
		\end{equation}where $\xi_k = 4J_1\big(2\tfrac{\delta}{\omega_{\rm d}}(g q^1_{\rm pn} + \lambda q^1_{\rm pt})\big)(gq^1_{\rm pn} + \lambda q^1_{\rm pt})\cos(k) \omega_{\rm d}/\delta$. For the initial conditions $n_{{\rm s},k}(0) = 0$ and $\dot{n}_{{\rm s},k}(0) = 0$, the solution can be readily obtained; however, by taking the limit $t \to \infty$ in the NESS, we arrive at:\begin{equation}\label{eq_17}
			n_{{\rm s},k}(t \to \infty) = {} \frac{\xi^2_k}{2(\gamma_{\rm s}^2 - \xi^2_k)} \, .
		\end{equation}Inserting this into Eq.~\eqref{eq_15}, for $v_{{\rm s},k}(t) = \Re[z_{{\rm s},k}(t)]$, we obtain\begin{equation}\label{eq_18}
			v_{{\rm s},k}(t \to \infty) = - \frac{\xi_k \gamma_{\rm s}}{2(\gamma_{\rm s}^2 - \xi^2_k)}\sin(h_k(t))\, .
		\end{equation}\\
		
		\hypertarget{mylinkC}{\blue{\textbf{\textit{Appendix C: Laser switch-off and off-resonance effects}}}}.---As the last additional information, let us see immediately what happens when the laser field is turned off after achieving the NESS~(e.g., at $t J= 1000$). In the context of the Lindblad quantum master equation, relaxation represents the system's return to thermal equilibrium after the driving field is removed. In our model in Fig.~\ref{f4}(a), the system is initially set to a temperature of zero when the drive is off, and it is verified that the system naturally relaxes back to this state once the drive is turned off again. Thus, a weak steady laser must remain on for a sufficient duration to ensure both magnetization switching and demagnetization processes occur, allowing the system to reach the NESS with certainty. \begin{figure}[b]
			\centering	\includegraphics[width=0.75\linewidth]{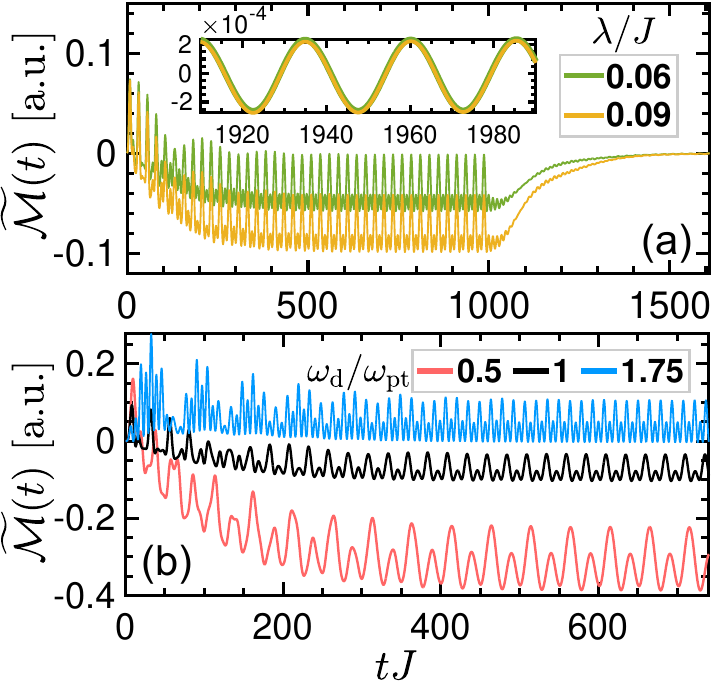}
			\caption{(a) Laser switch-off induces Néel magnetization relaxation to zero, enabling switching and demagnetization under varying spin–photon coupling $\lambda$. For resonant and redshift off-resonant drive (frequency $\leq$ photon frequency) switching and demagnetization occur, whereas for blueshift off-resonant drive ($>$ photon frequency) reduced magnon pair dominance suppresses these effects. Parameters: $\Delta = 0.0028\,J$, $\omega_{\rm pt} = \omega_{\rm pn} = J/4$, $g = J$, $\mathcal{E}_0 = J/400$, $\gamma_{\rm pn} = \gamma_{\rm pt} = \kappa = J/80$, and $\gamma_{\rm s} = J/100$). In (b), we set $\lambda = 9 J/100$.} 
			\label{f4}
		\end{figure}
		
		We also provide the effects of setting the laser frequency $\omega_{\rm d}$ off-resonance with the photon frequency $\omega_{\rm pt}$. Figure~\ref{f4}(b) presents two scenarios: redshifted (where $\omega_{\rm d} < \omega_{\rm pt}$) and blueshifted (where $\omega_{\rm d} > \omega_{\rm pt}$). Under off-resonance conditions, magnetization switching or demagnetization is achievable only for redshifted frequencies relative to the resonant case ($\omega_{\rm d} = \omega_{\rm pt}$). This is primarily attributed to the condensation phase of matter, where particles tend to occupy low-energy states at low temperatures. These findings highlight the critical role of resonance and redshifted optical processes in achieving the primary objectives of this work.
	}
\end{document}